**Title**

Alzheimer's Disease is Associated with Isotropic Ocular Enlargement

**Authors**

Shuyue Ma [1,2], Qihui Ye [1,2], Chufan Xiao [1,2], Haifei Guan [1,2], Zhicheng Du [1,2], Peiwu Qin [1,2]

**Authors' affiliation**

[1]Tsinghua-Berkeley Shenzhen Institute, Tsinghua Shenzhen International Graduate School, Tsinghua University, Shenzhen 518055, China

[2]Institute of Biopharmaceutics and Health Engineering, Tsinghua Shenzhen International Graduate School, Shenzhen, Guangdong Province, 518055, China.

Address for correspondence and reprints: pwqin@sz.tsinghua.edu.cn.

**ABSTRACT**

Recent studies have documented ocular changes in dementia patients, especially Alzheimer's Disease (AD). In this study, we explored the change of eye size and eye shape in dementia, including AD patients. The eyeball volume and diameters were estimated via T1-weighted brain magnetic resonance (MR) images in the OASIS-3 database which included 83 AD, 247 non-AD dementia and 336 normal-aging participants qualified for this study. After adjustment of age, sex, race, apolipoprotein E genotypes, anisotropic ratio and intracranial volume, we observed the eyeball volume of the AD group was significantly larger than both the normal control (6871mm$^3$ vs 6415mm$^3$, $p < 0.001$) and the non-AD dementia group (6871mm$^3$ vs 6391 mm$^3$, $p < 0.001$), but there was no difference between the non-AD dementia group and the normal control (6391 mm$^3$ vs 6415mm$^3$, $p = 0.795$). Similar results were observed for the axial, transverse and vertical length. No group differences were observed in the anisotropic ratio, indicating an isotropic volume increase consistent with previous changes induced by the ocular hypertension (OH), which suggested possible elevation of the intraocular pressure (IOP) in AD. In consideration of the recent findings in ocular changes of dementia, our findings emphasize routine eye examinations and eye cares for AD patients in the clinic.

## INTRODUCTION

Dementia is a syndrome characterized by the deterioration of cognitive function beyond normal aging [1]. It mainly affects the elderly, with a prevalence of 5.2% in people over the age of 60 globally [2]. The World Health Organization (WHO) estimated the number of people with dementia to be 55 million worldwide in 2023, with nearly 10 million new cases every year [3]. Dementia results from a variety of diseases or injuries, with Alzheimer's Disease (AD) the most common form, which happens in parallel with accumulating amyloid β (Aβ) and Tau pathology [4]. Besides the changes in the brain, ocular changes also exist in dementia, especially AD. Retinal nerve fiber layer (RNFL) thinning has been suggested as a preclinical biomarker of dementia [5, 6]. Aβ accumulation inside and around retinal ganglion cells (RGCs) are also found in postmortem AD eyes, providing another in vivo biomarker for early AD detection [7-9]. However, the morphological change of the eye could be an indicator of eye disorders such as myopia with an elongation of the axial length and ocular hyperopia (OH) which may lead to the increase of ocular volume [10, 11]. Myopia was suspected to be correlated with dementia, yet with equivocal evidences. Previous study reported an almost 2- fold increased likelihood of suffering cognitive dysfunction in myopic patients compared with normal vision controls after adjusted for covariates, while another study found no significant association [12-14]. The results could be confounded by the inaccurate evaluation of visual acuity (VA) in patients with dementia due to the impaired cognition [15]. Recent studies have adopted advanced imaging methods or deep learning approaches to assist the detection [16-21]. Separately, some discovered slightly elevated intraocular pressure (IOP) in the mouse model of AD with unknown significance[22], with others reported no significant differences between IOP of the AD patients with that of the control subjects [23]. Therefore, a deeper understanding of the eye size and eye shape may provide supporting information on ocular abnormalities in patients with dementia. In this study, we measured the volume and diameters of the eye via structural brain magnetic resonance (MR) imaging in the Open Access Series of Imaging Studies-3 (OASIS-3), a large clinical dataset on dementia. Hopefully, the evidences of ocular morphological changes could have the potential to improve regular eye cares for AD patients in the clinic.

## METHODS

### Dataset description

Data used in this article were obtained from the Open Access Series of Imaging Studies-3 (OASIS-3) database (http://oasis-brains.org). OASIS-3, collected by Washington University Knight Alzheimer Disease Research Center, provided MR imaging and related clinical data of 1474 participants, consisting of 755 cognitively normal adults and 719 individuals at various stages of cognitive decline. Other information in OASIS-3 database included apolipoprotein E genotypes (APOE), the main genetic determinants of Alzheimer disease (AD) and volumetric segmentations of T1-weighted images processed through an XNAT pipeline for the FreeSurfer image analysis suite (http://surfer.nmr.mgh.harvard.edu/).

### Participants

Participants were assessed through clinical protocols in accordance with National Alzheimer's Coordinating Center Uniform Data Set (UDS) [26]. For each participant, OASIS-3 documents his entries in a time series. Subjects were treated as normal controls (NCs) only if they were cognitively normal through all visits, while participants who met the criteria for dementia at least once were classified as the dementia group. Since the average age of subjects with dementia are significantly larger than controls, we extracted the visits when individuals were assessed as dementia for the first time and normal for the last time to balance the age between groups. For example, if a participant had 10 diagnostic records in total and was first diagnosed as dementia in the 5th time, we then extract the 5th record as the diagnostic data of this participant. The selected records were matched up with the closest T1w scanning with corresponding intracranial volume (ICV) if the time variation was within 3 years. Then participants who were not between 60 and 90 years old during entry were excluded, so did those with missing information on race or APOE genotype. During processing, images with low resolution or artifacts, and outliers with abnormal intracranial or ocular volume were also excluded. The final dataset used in this study contained 666 subjects, consisting of 330 subjects with dementia and 336 normal participants (Fig 1).

In clinical assessment, dementia was divided into different subtypes dependent on the family history, medical history, physical examination, and neurological evaluation [26]. Among the 330 participants in dementia group, there were 83 subjects diagnosed as 'AD', 175 as 'probable AD', 32 as 'possible AD', 3 as 'vascular dementia' and 3 as 'dementia with Lewy bodies'. In this study, we adopted the strict criteria for AD diagnosis by classifying "AD" as a separate subgroup and the rest as 'non-AD dementia'.

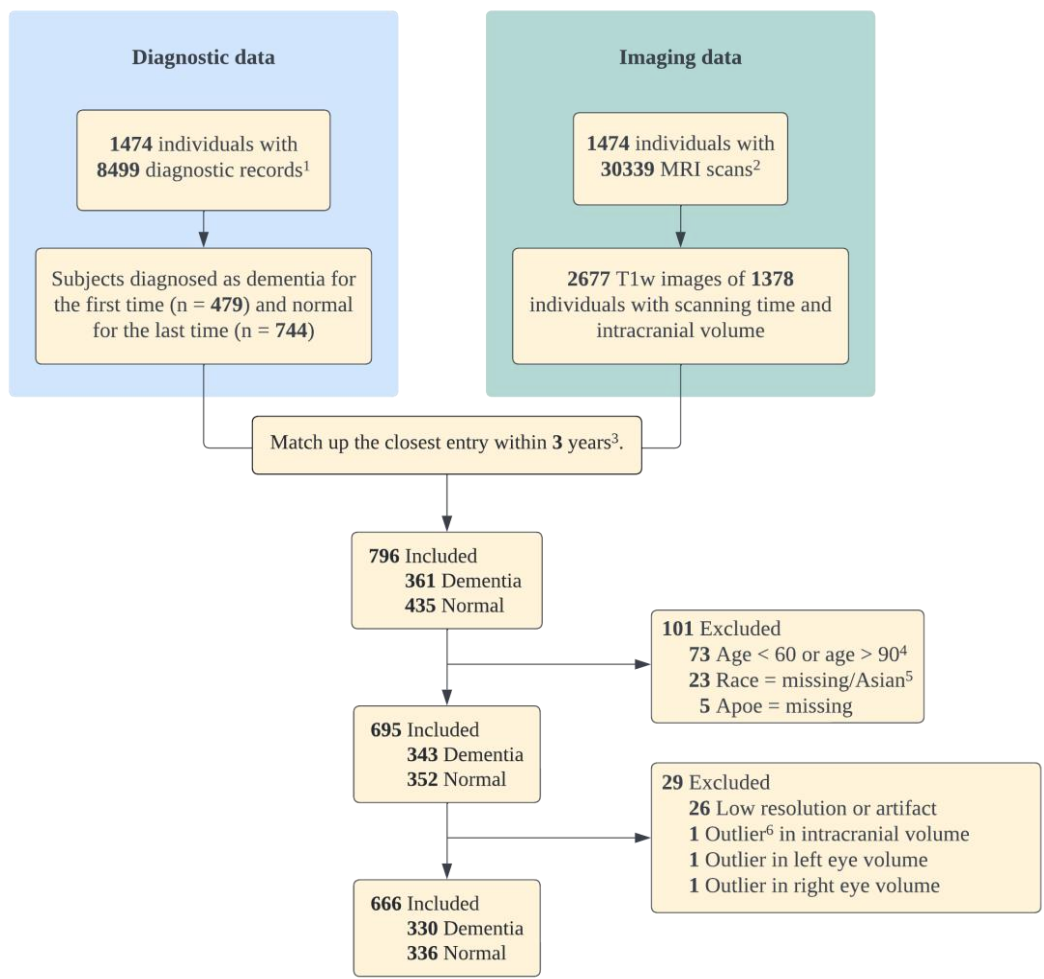

**Figure 1. Flow diagram of the subject selection procedure.** [1,2] For each subject, OASIS-3 provides records over time both in clinical diagnosis and MR imaging, so that the cogitive assessments of 1474 individuals are 8499 times, and the number of MRI scans are 30339 in total, yet with unmatched date. [3] If the time difference between diagnosis and scan of a participant is less than 3 years, they are integrated for further analysis. [4] The age discrepancy remains distinct due to the existence of upper ouliers in subjects with dementia and lower outliers in NC, so we restrict the age interval. [5] Subjects with an Asian race (n=11) are excluded for the simplicity of adjustment in a linear model. [6] An outlier is defined as locating outside 1.5 times the interquartile range above the upper quartile or below the lower quartile

**MRI processing and analysis**

Magnetic resonance imaging (MRI) used in this study were T1-weighted images, consistent with where ICV was obtained. All images were in nifti format and processed in 3D Slicer 5.2.1 (http://www.slicer.org).

We measured the lengths of diameters in the axial, coronal and sagittal plane. Concretely, a central point was marked by locating the center of eyeball in the largest coronal plane. Then the axial length was defined as the distance from the anterior corneal surface to the retinal membrane in the direction perpendicular to the lens diameter, while the transverse length was defined as the distance between temporal and nasal ends of the eye perpendicular to the axial length in the largest axial view. Similarly, vertical length was defined as the distance between top and bottom ends of the eye perpendicular to the axial length in the largest sagittal plane (Fig 2).

To estimate the ocular volume, the precise approach is 3D segmentation. By setting the intensity threshold, erasing non-target regions, and interpolating among slices, the eyeballs were segmented and labeled. The volume was then calculated by the number and size of segmented voxels in 3D Slicer. Despite taking account of the irregular shape of eyeballs, 3D segmentation is nonetheless time consuming. By considering the eye as an ellipsoid for approximation, 2D measurements were more efficient, where the estimated volume was calculated as:

$$Estimated\_volume = \frac{4}{3}\pi \cdot \frac{Axial\_length}{2} \cdot \frac{Transverse\_length}{2} \cdot \frac{Vertical\_length}{2}.$$

Since the relative size of both eyes were highly coincident and alternations always happen simultaneously, for a single individual, we toke the average volume of the left and right eye. We also assessed the linear correlation between 3D segmentation and 2D measurement to ensure that the latter was a proper substitution when assessing the ocular volume.

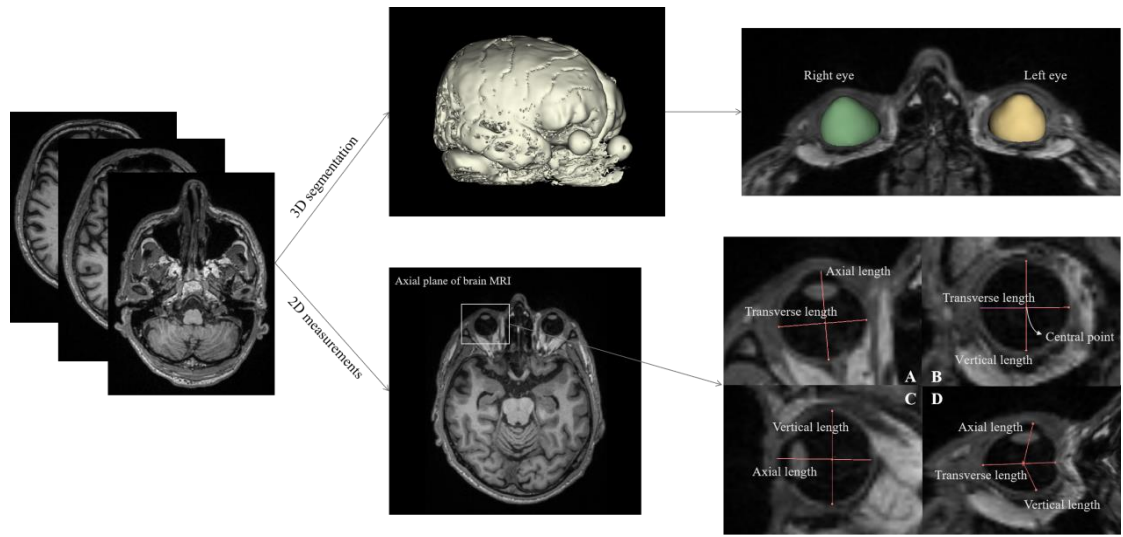

**Figure 2. 3D segmentation and 2D measurements to extract eyeball volume through brain MR images**. MRI slices were in nifti format and processed in 3D Slicer. (A) Axial plane; (B) Coronal plane; (C) Sagittal plane; (D) 3D view of the measurements.

## Statistical analysis

All analyses were performed in Stata, version 15. Descriptive statistics were presented as means, standard deviations for continuous variables and proportions for categorical variables. Linear regression with post-hoc multiple comparison was used to test the associations of dementia with the estimated eyeball lengths and volume. Age, sex, race and APOE4 genotypes were included as covariates. For the APOE, it has three variants in humans (ε2, ε3, and ε4) and ε4 is linked to an increased risk for AD [28], so we classified it into non e4 carrier (ε2/ε2, ε2/ε3, ε3/ε3), e4 carrier (ε2/ε4, ε3/ε4) and e4 homozygote (ε4/ε4). Furthermore, to confirm that volume differences were due to the isotropic enlargement, not merely the elongation of axial length, or a larger ICV, ICV and the anisotropic ratio (which is defined as: anisotropic_ratio = Axial_length / Transverse_length) were also included as covariates. Statistical significance was set at $P < 0.05$.

## RESULTS

Demographic data were summarized in Table 1. Both non-AD dementia and AD groups were older and had more females than the normal controls. The average ICV of non-AD dementia group was also larger than the normal controls. In comparison with the normal controls, he non-AD dementia group had more participants being e4 carriers, while the AD dementia group had more participants being e4 homozygotes.

**Table 1. General characteristics of the participants**

| Characteristic | Normal | Non-AD dementia | Alzheimer's |
| --- | --- | --- | --- |
| N | 336 | 247 | 83 |
| Age, y | 71.3 ± 6.6 | 76.5 ± 6.6** | 75.0 ± 6.3** |
| Female, No.(%) | 199 (59.2%) | 178 (72.1%)* | 62 (74.7%)* |
| Race | | | |
|   White | 278 (82.7%) | 211 (85.4%) | 75 (90.4%) |
|   Black | 58 (17.3%) | 33 (13.4%) | 8 (9.6%) |
| APOE |   |   |   |
|    Non e4 carrier | 217 (64.6%) | 104 (42.1%)** | 32 (38.6%)** |
|    e4 carrier | 113 (33.6%) | 129 (52.2%)** | 37 (44.6%) |
|    e4 homozygote | 6 (1.8%) | 14 (5.7%) | 14 (16.9%)** |
| Education |   |   |   |
|    High school or below | 42 (12.5%) | 53 (21.5%)* | 13 (15.7%) |
|    College or below | 179 (53.3%) | 130 (52.6%) | 37 (44.6%) |
|    Graduate school or above | 115 (34.2%) | 59 (23.9%) | 33 (39.8%) |
| Intracranial volume, mm$^3$ | 1484081 ± 172219.2 | 1560912 ± 180223.9** | 1520009 ± 149810 |

*Abbreviations: SD, standard deviation; AD, Alzheimer's disease; APOE, apolipoprotein E gene; ** $p < 0.001$, * $p < 0.01$ compared with normal group. 5 were missing in the education level of Non-AD dementia group.*

Among the 666 selected individuals, the eyeball diameter measurement was performed on all participants, while 3D segmentation was conducted on 250 individuals. As shown in Fig 3, there was a strong linear correlation between the segmented volume and the estimated volume ($r = 0.93$; $p < 0.001$), indicating that measuring the diameters could be used as an alternative approach to estimate the eyeball volume.

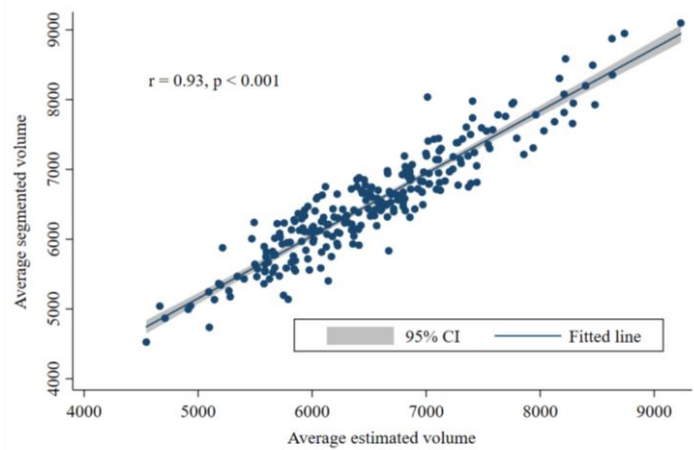

**Figure 3. Correlation between the segmented volume and estimated volume**

*Abbreviations: CI, confidence interval.*

After adjustment of age, sex, race, APOE4, anisotropic ratio and ICV, we found the eyeball volume of Alzheimer's group was significantly larger than both the normal control (6871mm$^3$ vs 6415mm$^3$, $p < 0.001$) and the non-AD dementia group (6871mm$^3$ vs 6391 mm$^3$, $p < 0.001$), but there was no difference between the non-AD dementia group and the normal control (6391 mm$^3$ vs 6415mm$^3$, $p = 0.795$) (Figure 4A). Similar results were observed for the axial, transverse and vertical length (Figure 4B-4D). No group differences were observed in the anisotropic ratio, indicating an isotropic volume increase in AD patients.

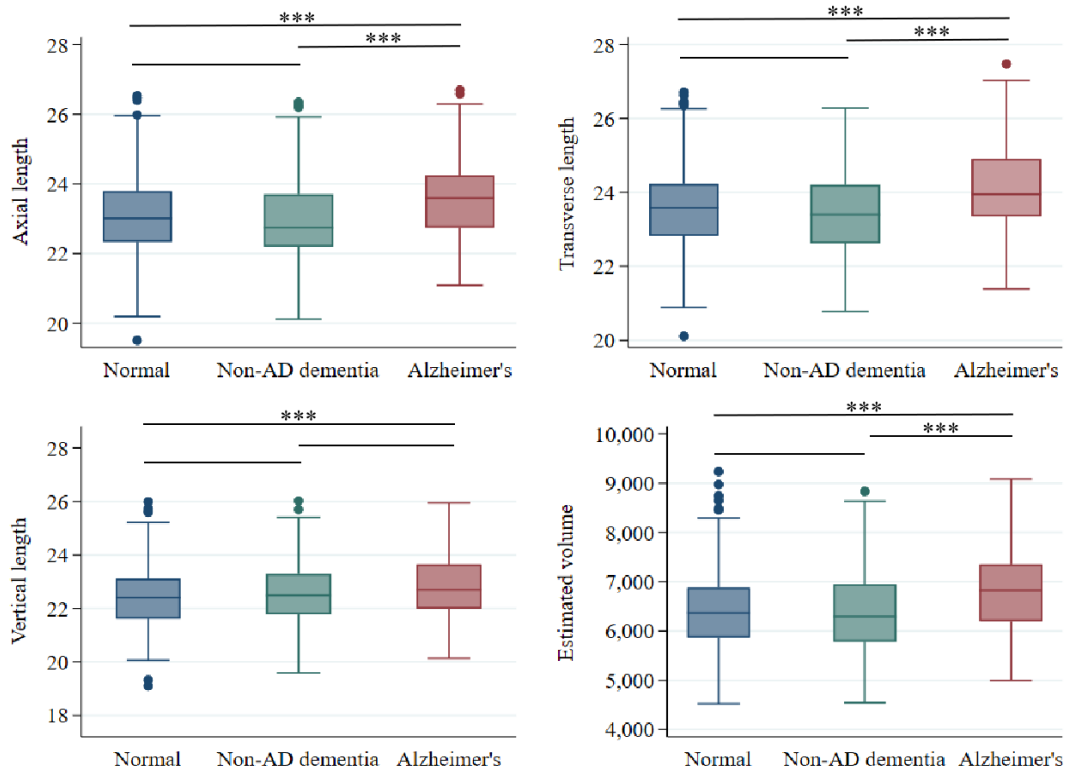

**Fig 4. Comparison of the eyeball length and volume in normal, non-AD dementia and Alzheimer's group.** *Abbreviations: AD, Alzheimer's disease; \*\*\* p < 0.001. P-values of the eyeball length were adjusted for age, sex, race and APOE4, while p-values of the ocular volume were adjusted for age, sex, race, APOE4, anisotropic ratio and ICV. In each box, from top to bottom, five lines represented upper adjacent value, 75th percentile, median, 25th percentile and lower adjacent value respectively. Dots represented outside values.*

## DISCUSSION

This is the first study to explore the ocular shape and volume changes in dementia. By using T1 weighted brain MR images, we compared the ocular volume and eyeball diameters among normal aging, non-AD dementia and Alzheimer's groups. We found a significant larger eyeball volume in AD but not in non-AD dementia, after adjusted for age, sex, race, APOE4, anisotropic ratio and ICV. Specifically, the diameter increase was significant in axial, transverse and vertical directions, indicating an isotropic enlargement of the eye, similar to previous findings on the glaucomatous eyeball change that caused by IOP increase [27]. Hence we suspect that the dimensional change of the eye in AD patients is more likely induced by the OH.

IOP is maintained by the aqueous humor (AH) dynamics. AH is secreted to the posterior chamber by ciliary epithelial cells and leaves the eye via two pathways, the trabecular meshwork (conventional) and uveoscleral (unconventional) pathway. Dysfunction in the balance between AH formation and outflow can lead to an increased IOP, as what commonly occurs in the glaucomatous eyes [29, 30]. Both adrenergic and cholinergic system play an important role in the AH flow pathway. NE was known to reduce IOP by both decreasing AH production and increasing AH outflow. However, NE is mainly supplied by the LC, tau deposition of which may be the first identifiable pathology of AD [31]. Postmortem studies indicated an 80% reduction of LC cell number in AD compared with age-matched controls [32]. The accumulated neuron loss in LC can lead to the reduction of NE concentration and abnormalities of noradrenergic innervation [33, 34]. It is worth noting that decreased NE level was not observed in all patients with cognitive decline. The LC neurodegeneration was found occurred at early stages and contributed to the progression of diseases only in AD and dementia with lewy bodies (DLB), but not in dementia with other contributing factors such as vascular dementia [35, 36]. In patients diagnosed as probable or possible AD, their symptoms were less typical and more likely caused by multiple pathogenetic factors. Therefore, the malfunction of LC may not serve as a primary contribution and may develop less dramatically with the disease progressing. This may explain why the ocular enlargement was not observed in 'probable AD', or 'possible AD' (6344mm$^3$ vs 6415mm$^3$, p = 0.66 and 6254mm$^3$ vs 6415mm$^3$, p = 0.55, respectively). Similar mechanisms were also found in the cholinergic system. Cholinergic signals, including ACh and muscarinic acetylcholine receptors (AChRs) activation, can cause the contraction of ciliary muscle and widening of anterior chamber angle, increasing aqueous humor outflow through unconventional pathway [37, 38], but were found to be dysregulated in AD [39]. Nevertheless, whether these signaling changes in AD enough to cause IOP alternation requires further studies.

AD is characterized by the deficient Aβ clearance of cerebrospinal fluid (CSF), resulting in an impaired "glymphatic" pathway, a glial cell-dependent perivascular network clearing metabolic waste from the parenchyma [40, 41], Recent studies documented the presence of an ocular glymphatic system, where CSF entered the ON parenchyma and drained into meningeal lymphatic vessels [42, 43]. Dysfunction of ocular glymphatic system and the subsequent accumulation of Aβ can induce significant RGCs apoptosis, ultimately leads to vision damage [8, 44]. Moreover, this process was pressure dependent. An elevated translaminar cribrosa pressure difference (TLCPD), defined as IOP minus CSF pressure (CSFP), impaired the intra-axonal clearance and increased the risk of optic neuropathy [45, 46]. The fact that AD patients were previously found to have a low CSFP [47], in combination with the finding of this study, i.e. a potential high IOP in AD patients, will aggravate the burden to ocular glymphatic pathway. Therefore, this study suggests that regular eye screenings, including the visual acuity, IOP test and fundoscopic examination are necessary for AD patients in clinic [48].

The study also has limitations. First, this is the first study to explore morphological differences of the eye in patients with dementia and the conclusion is drawn on a unitary dataset, which need to be replicated by future studies. Second, this is a cross-sectional data. Future longitudinal studies are warranted to explain the nature and direction of causality. Third, this study lack explicit ocular indicators, especially the measurement of IOP. Subjects suspected to have dementia are usually not arranged to take IOP test in the clinic, therefore, the existing large datasets on dementia do not have IOP from the tonometry test. Future replication studies need to add tonometry test into clinical exams for participants in dementia study.

In conclusion, our study observed that patients with AD dementia had a significant larger eyeball volume, similar to the change induced by OH. In consideration of the recent findings in ocular changes of dementia, our findings emphasis more regular and routine eye examinations and eye cares for AD patients in the clinic.

## ACKNOWLEDGEMENTS

Data were provided by OASIS-3 Principal Investigators: T. Benzinger, D. Marcus, J. Morris; NIH P50AG00561, P30NS09857781, P01AG026276, P01AG003991, R01AG043434, R01AG054567, UL1TR000448, and R01EB009352.

## CONFLICT OF INTEREST

There is no competing interests or potential conflicts in this study.